\documentclass{PoS}

\title{Chiral symmetry restoration and eigenvalue density of Dirac operator at finite temperature }

\ShortTitle{Chiral symmetry restoration and eigenvalue density of Dirac operator}

\author{\speaker{Sinya Aoki}\\
    Graduate School of Pure and Applied Sciences, University of Tsukuba, Tsukuba, Ibaraki 305-8571, Japan, and  \\
    Center for Computational Sciences, University of Tsukuba, Tsukuba, Ibaraki 305-8577, Japan \\ 
        E-mail: \email{saoki@het.ph.tsukuba.ac.jp}}

\author{Hidenori Fukaya\\
        Department of Physics, Osaka University, Toyonaka 560-0043, Japan\\
        E-mail: \email{hfukaya@het.phys.sci.osaka-u.ac.jp}}

\author{Yusuke Taniguchi\\
Graduate School of Pure and Applied Sciences, University of Tsukuba, Tsukuba, Ibaraki 305-8571, Japan, and  \\
    Center for Computational Sciences, University of Tsukuba, Tsukuba, Ibaraki 305-8577, Japan \\ 
        E-mail: \email{tanigchi@het.ph.tsukuba.ac.jp}}
        
\abstract{We reinvestigate constraints on the eigenvalue density of the Dirac operator in the chiral symmetric phase of 2 flavor QCD at finite temperature, employing the overlap Dirac operator with
 the exact chiral symmetry at finite lattice spacings 
to avoid possible ultra-violet(UV) divergences. 
Studying multi-point correlation functions in various channels in
the thermodynamical limit, 
we obtain stronger constraints than those found in the previous studies that
not only the eigenvalue density at the origin but also its first and second derivatives vanish in the chiral limit of 2 flavor QCD.
In addition  we show that the axial U(1) anomaly becomes invisible in susceptibilities of scalar and pseudo scalar mesons.  }

\FullConference{The 30 International Symposium on Lattice Field Theory - Lattice 2012,\\
		June 24-29, 2012\\
		Cairns, Australia}

\begin{document}

\section{Introduction}
While the classical QCD Lagrangian with $N_f$ massless quarks poses
SU$(N_f)_L\times$ SU$(N_f)_R\times$ U$(1)_V \times$ U$(1)_A$ chiral symmetry,
U$(1)_A$ part is broken explicitly at quantum level by the anomaly and then
SU$(N_f)_L\times$ SU$(N_f)_R$ part is spontaneously broken to SU$(N_f)_V$ in the QCD vacuum.
On the other hand,  it is expected that the SU$(N_f)_L\times$ SU$(N_f)_R$ chiral symmetry is recovered above the critical temperature $T_c$, and 
and the first principle lattice QCD calculations support this expectation.
It still remains an open question, however, if, how and when the U$(1)_A$ part is restored.

The question whether the U$(1)_A$ symmetry is restored or not near $T_c$ is of phenomenological importance\cite{Pisarski:1983ms}. Furthermore a connection between the restoration of U$(1)_A$ symmetry and the gap in the eigenvalue density of Dirac operator near the origin is suggested\cite{Cohen:1997hz}.
In principle, the fate of U$(1)_A$ symmetry can be investigated by numerical lattice QCD simulations.
On going four simulations with different quark actions, however, have reported different results\cite{Ohno:2011yr,Bazavov:2012ja,Kovacs:2011km,Cossu:2012gm}.
Analytical investigations in the continuum QCD are also inconclusive\cite{Cohen:1996ng,Lee:1996zy}

In this report, we address these problems on a lattice, but using an analytic method.
We concentrate on $N_f=2$ case and employ the overlap Dirac operator\cite{Neuberger:1997fp,Neuberger:1998wv}, which ensures the exact chiral symmetry\cite{Luscher:1998pqa} through the Ginsparg-Wilson (GW) relation\cite{Ginsparg:1981bj} while breaks the U$(1)_A$ symmetry by the fermionic measure\cite{Hasenfratz:1998ri}. Using the spectral decomposition of correlation functions and assuming the restoration of the non-singlet chiral symmetry, we extract new constraints on the Dirac eigenvalue density in addition to the manifest one implied by the Banks-Casher relation\cite{Banks:1979yr}.  
We also discuss the fate of U$(1)_A$ symmetry at $T\ge T_c$ using these constraints.

In Sec.~\ref{sec:setup}, we explain our setup and assumptions of our analysis.
Our main results, constraints on the eigenvalue density, are given in Sec.~\ref{sec:result}.
The fate of U$(1)_A$ symmetry is discussed in Sec.~\ref{sec:singlet}.
For more details of our analysis and results, see Ref.~\cite{AFT}.

\section{Setup}
\label{sec:setup}
\subsection{Spectral decomposition of the overlap fermions}
The quark part of $N_f$-flavor lattice QCD action is given by
\begin{eqnarray}
S_F &=& a^4\sum_x\left[ \bar\psi D(A)\psi + m \bar \psi F ( D(A) ) \psi \right](x), \quad
F(D) = 1- \frac{ R a}{2} D,
\end{eqnarray}
where $\psi =(\psi_1,\psi_2,\cdots,\psi_{N_f} )^T$ denotes the set of $N_f$ quark fields with the degenerate mass $m$, $a$ is the lattice spacing, and $D(A)$ is the overlap Dirac operator\cite{Neuberger:1997fp,Neuberger:1998wv} for a given gauge configuration $A$, which satisfies the GW relation\cite{Ginsparg:1981bj} that
\begin{eqnarray}
D(A)\gamma_5 +\gamma_5 D(A) = a D(A) R \gamma_5 D(A) , 
\end{eqnarray}
with an arbitrary constant $R$, and the $\gamma_5$ hermiticity that $D(A)^\dagger = \gamma_5 D(A)\gamma_5$.

Let us consider eigenvalues and eigenfunctions of $D(A)$, $D(A) \phi_n^A = \lambda_n^A\phi_n^A$.
The GW relation implies that
\begin{eqnarray}
\lambda_n^A + \bar\lambda_n^A &=& a R \bar\lambda_n^A \lambda_n^A,\quad
D(A)\gamma_5 \phi_n^A = \bar\lambda_n^A\gamma_5\phi_n^A,
\end{eqnarray}
where $\bar \lambda_n^A$ is a complex conjugate of $\lambda_n^A$, and $\gamma_5 \phi_n^A$ is the corresponding eigenfunction. With an inner product defined by $(f, g) = a^4\sum_x f^\dagger(x) g(x) $,
eigenfunctions with complex eigenvalues can be ortho-normal as $(\phi_n^A,\phi_m^A) = (\gamma_5\phi_n^A,\gamma_5\phi_m^A)=\delta_{nm}$ and $(\phi_n^A,\gamma_5 \phi_m^A) = 0$.
For the real eigenvalues, $\lambda_k^A =0$ (zero modes) and $\lambda_K^A = 2/(Ra)$ (doubler modes), corresponding eigenfunctions
can be made chiral, so that we denote the number of left(right)-handed zero modes as $N^A_L (N^A_R)$ and that of doubler modes as $n^A_L (n^A_R)$. 

The propagator of the massive overlap quark for each flavor can be expressed in terms of these eigenvalues and eigenfunctions as
\begin{eqnarray}
S_A(x,y) &=& \sum_{\{{\rm Im}\, \lambda_n^A > 0\}}
\left[\frac{\phi_n^A(x) \phi_n^A(y)^\dagger}{f_m \lambda_n^A+ m} +
\frac{\gamma_5 \phi_n^A(x) \phi_n^A(y)^\dagger\gamma_5}{f_m \bar\lambda_n^A+ m}
\right] \nonumber \\
&+&\sum_{k=1}^{N_{R+L}^A}\frac{\phi_k^A(x)\phi_k^A(y)^\dagger}{m}
+\sum_{K=1}^{n_{R+L}^A}\frac{\phi_K^A(x)\phi_K^A(y)^\dagger}{2/(Ra)},
\label{eq:prop}
\end{eqnarray}
where $f_m = 1 - R ma/2$, $N_{R+L}^A =N_R^A+N_L^A$ and $n_{R+L}^A =n_R^A+n_L^A$.
A measure for a given gauge field $A$ can be also written in terms of eigenvalues as
\begin{eqnarray}
P_m(A) &=& e^{-S_{YM}(A)}\, m^{N_f N_{R+L}^A}\, \Lambda_R^{N_f n_{R+L}^A}\,
\prod_{\{{\rm Im} \lambda_n^A > 0\}} \left( Z_m^2 \bar\lambda_n^A\lambda_n^A + m^2\right)^{N_f},
\end{eqnarray}
where $S_{YM}(A)$ is the gauge part of the action, $\Lambda_R = 2/(Ra)$ and $Z_m^2 = 1 - m^2/\Lambda_R^2$. For even $N_f$, $P_m(A)$ is positive semi-definite and even function of $m$.
In this talk, we consider the $N_f=2$ case.

It is important to note that all quantities consist of $S_A(x,y)$ and $P_m(A)$ are finite at finite volume ($V < \infty$), the nonzero quark mass ($m\not=0$) and the finite lattice spacing ($a\not=0$).
We then carefully take the $V\rightarrow \infty$ limit and then $m\rightarrow 0$ limit, without worrying about possible ultra-violet divergences, until the continuum limit is taken. 

\subsection{Chiral Ward-Takahashi identities on the lattice}
With the GW relation, the lattice quark action at $m=0$  is invariant under lattice chiral rotation\cite{Luscher:1998pqa} that
\begin{eqnarray}
\delta_a\psi(x) = i\theta T_a \gamma_5 \left[1- Ra D(A)\right]\psi (x),\quad
\delta_a\bar\psi(x) = i\theta \bar\psi(x) T_a\gamma_5, 
\end{eqnarray}
where $\theta$ is an infinitesimal real parameter, and $T_a$  for $a=1,2,\cdots, N_f^2-1$ denotes the generator of SU$(N_f)$, and $T_0={\bf 1}_{N_f\times N_f}$ for U$(1)_A$. 

For the volume-integrals of scalar and pseudo scalar density operators defined by
\begin{eqnarray}
S_a &=& a^4 \sum_x \left[ \bar\psi T_a F(D(A))\psi \right](x),\quad
P_a = a^4 \sum_x \left[ \bar\psi T_a \gamma_5 F(D(A))\psi \right](x),
\end{eqnarray}
it is easy to show for $N_f=2$ that 
\begin{eqnarray}
\delta_b S_a &=& 2\delta_{ab} P_0, \quad \delta_b P_a = -2\delta_{ab} S_0,
\qquad ({\rm for}\ a,b =1,2,3 ), \\
\delta_0 S_a &=& \delta_a S_0 = 2 P_a,\quad \delta_0 P_a = \delta_a P_0 = -2 S_a,\qquad
 ({\rm for}\ a =0,1,2,3 ) .
\end{eqnarray}
If the chiral symmetry is restored, 
a product of these operators, 
$
{\cal O}_{n_1,n_2,n_3,n_4} = P_a^{n_1} S_a^{n_2} P_0^{n_3} S_0^{n_4}
$
with $a=1,2,3$, satisfies the chiral Ward-Takahashi (WT) identities that
\begin{eqnarray}
\lim_{m\rightarrow 0}\lim_{V\rightarrow\infty}\frac{1}{V^l}\langle \delta_a{\cal O}_{n_1,n_2,n_3,n_4} \rangle_m &=& - \delta_{a0}\lim_{m\rightarrow 0}\lim_{V\rightarrow\infty}\frac{1}{V^l} \langle J_0 {\cal O}_{n_1,n_2,n_3,n_4} \rangle_m 
\end{eqnarray}
where
\begin{eqnarray}
\langle {\cal O}(A)\rangle_m &\equiv& \frac{1}{Z}\int {\cal D}A\, P_m(A)\, {\cal O}(A), \quad
Z=\int {\cal D}A\, P_m(A), \\
J_0 &=& -2i N_f a^4\sum_x \sum_{N=n,k,K} \phi_N^A(x)^\dagger \gamma_5 \left( 1-\frac{R}{2}aD\right)\phi_N^A(x) = -2iN_f Q(A), 
\end{eqnarray}
$Q(A) = N_R^A - N_L^A$ is the index of the overlap Dirac operator\cite{Hasenfratz:1998ri} for a given gauge configuration $A$,
and $l$ is a minimum integer which makes the $V\rightarrow\infty$ limit finite in the above equation.
Here $J_0$ represents the effect of the U$(1)_A$ anomaly. 

\subsection{Basic properties and assumptions}
The eigenvalue density for a given gauge configuration $A$ is defined by
\begin{eqnarray}
\rho^A(\lambda) &=& \lim_{V\rightarrow\infty}\frac{1}{V}\sum_{\{{\rm Im}\, \lambda_n^A > 0\}}\delta\left(\lambda - \sqrt{\bar\lambda_n^A\lambda_n^A}\right),
\label{eq:expansion}
\end{eqnarray}
which is insensitive to the temperature $T$, since $T$ is fully controlled by $P_m(A)$. 
Note that $\rho^A(\lambda)$ is positive semi-definite for an arbitrary value of  $\lambda$ for all $A$. 
Although the original eigenvalue spectrum at finite $V$ is a sum of delta functions, we expect in the $V\rightarrow\infty$ limit that $\rho^A(\lambda)$ becomes a smooth function. We further assume that 
$\rho^A(\lambda)$ can be analytically expanded around $\lambda = 0$ that $\rho^A(\lambda) =\sum_{n=0}^\infty \rho_n^A \lambda^n/ n! $ within an arbitrary small convergence radius $\epsilon$. 
More precisely we here assume that configurations which do not have the expansion (\ref{eq:expansion}) are measure zero in the path integral with $P_m(A)$.

In the following analysis, we assume that the vacuum expectation values of $m$-independent function ${\cal O}(A)$ is analytic function of $m^2$ in the chiral symmetric phase.
Under this assumption, if
$\lim_{m\rightarrow 0} \langle {\cal O}(A)^{l_0}\rangle / m^k = 0$ for a non-negative integer $k$ and a positive integer $l_0$ for an  $m$-independent and positive semi-definite function ${\cal O}(A)$,
we can write
\begin{eqnarray}
\langle {\cal O}(A)^{l_0}\rangle_m &=& m^{2([k/2]+1)}\int {\cal D}A \hat P(m^2,A) {\cal O}(A)^{l_0}, 
\end{eqnarray}
where $[r]$ is the largest integer not larger than $r$, $\hat P(0,A) > 0$ at least for some $A$.
In other words, the leading $m$ dependence arises from the contribution of configurations which satisfy $\hat P(0,A) > 0$.
Using this expression, it is easy to show
\begin{eqnarray}
\langle {\cal O}(A)^{l}\rangle_m &=& m^{2([k/2]+1)}\int {\cal D}A \hat P(m^2,A) {\cal O}(A)^{l}
= O( m^{2([k/2]+1)} )
\end{eqnarray}
for an arbitrary positive integer $l$, since ${\cal O}(A)^{l_0}$ and ${\cal O}(A)^l$ are both positive and therefore share the same support in the configuration space.

\section{Constraints on eigenvalue density}
\label{sec:result}
In this section, we derive constraints on the eigenvalue density of the Dirac operator in the SU$(2)_L\times$
SU$(2)_R$ chiral symmetric phase. In this case, the non-singlet WT identities are written as
\begin{eqnarray}
\lim_{m\rightarrow 0}\lim_{V\rightarrow\infty}\frac{1}{V^l} \langle \delta_a{\cal O}_{n_1,n_2,n_3,n_4}\rangle_m &=& 0, \quad 
\quad {\cal O}_{n_1,n_2,n_3,n_4}\in {\cal O}_a^{(N)}
\end{eqnarray}
for $a\not=0$, where
\begin{eqnarray}
{\cal O}_a^{(N)} &\equiv& \left\{
{\cal O}_{n_1,n_2,n_3,n_4}\left\vert n_1+n_2 = \mbox{ odd }, \ n_1+n_3 = \mbox{ odd }, \
\sum_i n_i = N \right.
\right\}
\end{eqnarray}
is a set of non-singlet and parity odd operators of degree $N$. Explicitly we have
\begin{eqnarray}
\frac{\delta_a}{2}{\cal O}_{n_1,n_2,n_3,n_4} &=& -n_1{\cal O}_{n_1-1,n_2,n_3,n_4+1} + n_2{\cal O}_{n_1,n_2-1,n_3+1,n_4}-n_3{\cal O}_{n_1,n_2+1,n_3-1,n_4}+n_4{\cal O}_{n_1+1,n_2,n_3,n_4-1}.    
\nonumber
\end{eqnarray}

\subsection{Constraints at $N=1$}
At $N=1$, there is only one operator ${\cal O}_{1000} = P_a$ in ${\cal O}_a^{(N=1)}$, which gives
$\delta_aP_a = - 2S_0$. Using the decomposition in Eq.~(\ref{eq:prop}), we have
\begin{eqnarray}
\lim_{V\rightarrow\infty} \frac{1}{V}\langle -S_0\rangle_m &=&
\lim_{V\rightarrow\infty} \frac{N_f}{mV}\langle N_{R+L}^A\rangle_m + N_f \langle I_1\rangle_m,
\end{eqnarray}
where
\begin{eqnarray}
I_1 &=& \int_0^{\Lambda_R}\, d\lambda\, \rho^A(\lambda) \frac{ 2m g_0(\lambda^2)}{Z_m^2\lambda^2+m^2}=\pi \rho_0^A + O(m), \qquad g_0(x) = 1-\frac{x}{\Lambda_R^2} 
\end{eqnarray}
for small $m$.  The WT identity now becomes
\begin{eqnarray}
\lim_{m\rightarrow 0}\lim_{V\rightarrow\infty}\frac{1}{V}\langle -S_0\rangle_m &=&
\lim_{m\rightarrow 0}\lim_{V\rightarrow\infty}\frac{N_f}{mV}\langle N_{R+L}^A\rangle_m
+N_f \lim_{m\rightarrow 0} \langle I_1\rangle_m = 0
\end{eqnarray}
Since both $N_{R+L}^A$ and $I_1$ are positive,  the above equation gives two constraints that
\begin{eqnarray}
\lim_{V\rightarrow\infty}\frac{N_f}{mV}\langle N_{R+L}^A\rangle_m =O(m^2), \quad
\langle \rho_0^A\rangle_m = O(m^2) .
\end{eqnarray}

\subsection{Contribution from zero modes at general $N$}
Before considering $N=2,3,4$ cases in detail, we discuss the fate of zero mode contributions using WT identities at general $N$. Let us consider ${\cal O}_{1,0,0,N-1}\in {\cal O}_a^{(N)}$. The dominant term in  the non-singlet chiral WT identity in the large volume limit is given by
\begin{eqnarray}
-\frac{1}{V^N} \langle S_0^N \rangle = -(-1)^N N_f^N\left\langle \left(\frac{N_{R+L}^A}{mV}+I_1\right)^N\right\rangle_m +O(V^{-1}) .
\end{eqnarray}
Therefore,  the positivity of $N_{R+L}^A$ and $I_1$ leads to
\begin{eqnarray}
\lim_{V\rightarrow\infty} \frac{\langle ( N_{R+L}^A )^N\rangle_m}{V^N} &=& O\left( m^{2[N/2]+2}\right).
\end{eqnarray}

Since this holds for an arbitrary $N$, and $N_{R+L}^A$ does not explicitly depend on $m$, we conclude that
$\displaystyle \lim_{V\rightarrow\infty} \langle N_{R+L}^A\rangle_m/ V = 0$ at small but non-zero $m$. 

\subsection{Constraints at $N=2$}
At $N=2$ there are two conditions from the WT identities, 
\begin{eqnarray}
\chi^{\sigma-\pi} &=& \frac{1}{V^2}\langle S_0^2 - P_a^2\rangle_m \rightarrow 0, \quad
\chi^{\eta-\delta} = \frac{1}{V^2}\langle S_0^2 - P_a^2\rangle_m \rightarrow 0 ,
\end{eqnarray}
the former of which has already been examined in the previous subsection.
In terms of eigenvalues, we have
\begin{eqnarray}
\lim_{V\rightarrow\infty} \chi^{\eta-\delta} &=&\lim_{V\rightarrow\infty}\left\langle -\frac{N_f^2}{m^2V}Q(A)^2\right\rangle_m + N_f \left\langle \frac{I_1}{m} + I_2\right\rangle_m, 
\end{eqnarray}
where $I_2$ is defined by
\begin{eqnarray}
I_2 &=& 2\int_0^{\Lambda_R} d\lambda\, \rho^A(\lambda)\frac{m^2g_0^2(\lambda^2)-\lambda^2g_0(\lambda^2)}{(Z_m^2\lambda^2+m^2)^2}
\simeq\left(\frac{2}{\epsilon} +\frac{2\epsilon}{\Lambda_R^2}\right)\rho_0^A
+\left(2+\frac{\epsilon^2}{\Lambda_R^2}-\log\frac{\epsilon^2}{m^2}\right)\rho_1^A +O(1) .
\nonumber
\end{eqnarray}
Using 
\begin{eqnarray}
\frac{I_1}{m} + I_2 &=& \rho_0^A\frac{\pi}{m} + 2\rho_1^A +O(m), 
\end{eqnarray}
and $\langle \rho_0^A\rangle_m = O(m^2)$, the condition that
$\lim_{m\rightarrow 0}\chi^{\eta-\delta} = 0$ implies
\begin{eqnarray}
\lim_{V\rightarrow \infty} \frac{N_f \langle Q(A)^2\rangle_m}{m^2 V} = 2\langle \rho_1^A\rangle_m + O(m^2) .
\end{eqnarray}

\subsection{Constraints at $N=3,4$}
Five independent WT identities at $N=3$ give constraints that
\begin{eqnarray}
\langle \rho_0^A\rangle_m = - \frac{m^2}{2}\langle \rho_2^A\rangle_m + O(m^4), \quad
\lim_{V\rightarrow\infty}\frac{\langle Q(A)^2\rho_0^A\rangle_m}{m^2 V} = O(m^2) . 
\end{eqnarray}
Note that the second condition does not necessary give stronger constraint than $\langle Q(A)^2\rangle_m =O(m^2 V)$ and $\langle \rho_0^A\rangle_m =O(m^2 )$, since it only requires that a set of gauge configuration which satisfy both $Q(A)^2=O(V)$ and $\rho_0^A= O(1)$ has a weight proportional to $m^4$. 

After a little complicated manipulation, WT identities at $N=4$ lead to\cite{AFT} 
\begin{eqnarray}
\langle \rho_0^A\rangle_m &=& O(m^4), \quad
\langle \rho_1^A\rangle_m = O(m^2), \quad
\langle \rho_2^A\rangle_m = O(m^2), \quad
\lim_{V\rightarrow \infty} \frac{\langle Q(A)^2\rangle_m}{ V} =O(m^6) .
\end{eqnarray}

\subsection{Final results}
We finally obtain
\begin{eqnarray}
\lim_{m\rightarrow 0} \langle \rho^A(\lambda)\rangle_m &=& \langle \rho^A_3\rangle_0
\frac{\lambda^3}{3!} + O(\lambda^4).
\end{eqnarray}
We think that this condition is  the strongest since we know that the $N_f=2$ massless free quark theory has non-zero $\langle \rho_3^A\rangle_0$ keeping the exact non-singlet chiral symmetry.
Constraints at $N= 4 k$ ($k=1,2,\cdots$) in addition lead to $\langle \rho_0^A\rangle_m = 0$ at a small but non-zero $m$.

For the discrete zero modes, we have obtained
\begin{eqnarray}
\lim_{V\rightarrow\infty} \frac{1}{V}\langle N_{R+L}^A\rangle_m &=& 0, \qquad
\lim_{V\rightarrow\infty} \frac{1}{V}\langle Q(A)^2\rangle_m = 0
\label{eq:zeromode}
\end{eqnarray}
at a small but non-zero $m$, so that these zero-modes give no contribution to susceptibilities in the WT identities. 

\section{Discussions: singlet susceptibilities}
\label{sec:singlet}
For the results in the previous section, the singlet WT identity at $N=2$ vanishes as
\begin{eqnarray}
\chi^{\pi-\eta} &=& \frac{1}{V^2}\langle P_a^2-P_0^2\rangle_m = \lim_{V\rightarrow \infty}\frac{N_f^2}{m^2V}\langle Q(A)^2\rangle_m = 0
\end{eqnarray}
for small but non-zero $m$, if the non-singlet chiral symmetry is restored at $T \ge T_c$. 

For more general cases, the singlet WT identities is written as
\begin{eqnarray}
\lim_{m\rightarrow 0}\lim_{V\rightarrow\infty}\frac{1}{V^l}\langle \delta_0 {\cal O} \rangle_m &=& 2iN_f\lim_{m\rightarrow 0}\lim_{V\rightarrow\infty}\frac{1}{V^l}\langle Q(A) {\cal O} \rangle_m
\end{eqnarray}
where $l$ is the minimum integer which makes the $V\rightarrow\infty$ limit finite.
After a little algebra, the right-hand side of the above equation becomes\cite{AFT}
\begin{eqnarray}
\lim_{V\rightarrow\infty}\frac{1}{V^l}\langle Q(A) {\cal O} \rangle_m
&=& \lim_{V\rightarrow\infty} \left\langle \frac{Q(A)^2}{mV}\times  O(V^0) \right\rangle_m
=0
\end{eqnarray}
at small but non-zero $m$, thanks to Eq.~(\ref{eq:zeromode}). We therefore conclude that, for a class of operators considered in this report, the U$(1)_A$ breaking effect is invisible in the $V\rightarrow \infty$ limit.
This result suggests that the chiral phase transition for 2 flavor QCD  is likely to be of first order\cite{Pisarski:1983ms},
contrary to the naive expectation that the transition belongs to the O(4) universality class.

\vspace{0.2cm}

This work is supported in part by the Grant-in-Aid of MEXT (No. 22540265), the Grant-in-Aid for Scientific Research on Innovative Areas (No. 2004: 20105001, 20105003, 23105701,23105710) and SPIRE (Strategic Program for Innovative Research).

\end{document}